\documentclass[showpacs,preprintnumbers,amsmath,amssymb,aps,twocolumn,superscriptaddress]{revtex4}

\usepackage{amsfonts}
\usepackage[dvips]{graphicx} 
\usepackage{type1cm}
\usepackage{color}
\usepackage{bm}
\usepackage{epstopdf}

\newcommand{\figref}[1]{Fig.~\ref{#1}}

\def\be{\begin{equation}}
\def\ee{\end{equation}}
\def\bea{\begin{eqnarray}}
\def\eea{\end{eqnarray}}

\begin{document}
\title{An Introduction to Gauge Gravity Duality and Its Application in Condensed Matter}
\author{A.G.~Green}
\email{andrew.green@ucl.ac.uk}
\affiliation{London Centre for Nanotechnology, University College London, 17-19 Gordon St, London, WC1H 0AH, UK}

\begin{abstract} 

\end{abstract}

\maketitle

\vspace{0.2in}
{\bf
The past few years have witnessed a remarkable crossover of string theoretical ideas from the abstract world of geometrical forms to the concrete experimental realm of  condensed matter physics\cite{ADSCMTBook,HartnollReview,Lee:2010fk,McGreevyReview,Hartnoll:2011wl}. The basis for this --- variously known as holography, the AdS/CFT correspondence\footnote{Or distressingly for the abbreviation averse and mildly dyslexic amongst us, AdS/CMT.} or gauge-gravity duality --comes from notions right at the cutting edge of string theory. Nevertheless, the insights afforded can often be expressed in ways very familiar to condensed matter physicists, such as relationships between response functions and new sum rules.

The aim of this short, introductory review is to survey the ideas underpinning this crossover, in a way that -- as far as possible -- strips them of sophisticated mathematical formalism, whilst at the same time retaining their fundamental essence. I will sketch the areas in which progress has been made to date and highlight where the challenges and open questions lie. 
Finally, I will attempt to give a perspective upon these ideas. What contribution can we realistically expect from this approach and how might it be accommodated into the canon of condensed matter theory? Inevitably, any attempt to do this in such a rapidly evolving field will be superseded by events. Nevertheless, I hope that this will provide a useful way to think about gauge-gravity duality and the uncharted directions in which it might take us. }

\section{Introduction}

\subsection{ Duality in Condensed Matter Physics}
A new perspective often brings unexpected insights. In theoretical physics such changes of perspective are frequently afforded by transformations to new variables - related to the original variables by some (possibly complicated) mathematical transformation - in terms of which the underlying physics is more transparent. At its most simplistic level, the change of variable can make a mathematical problem easier to solve. A change of concept associated with this change of variable, however,  can have a more dramatic effect upon our understanding.
Such dualities are  common 
in condensed matter physics.
 Examples include changing from a description of the local deviation of the positions of the atoms in a crystal to one in terms of extended dislocations -- a classical example of a gauge-string duality. Indeed, motion of an electron in a crystal that contains a finite density of dislocations can be mapped very closely to motion in a gravitational field.

The study of strongly correlated quantum systems has  thrown up many cases where the most useful description is in term of gauge degrees of freedom. These are of particular relevance to the application of gauge-gravity duality.
The connection of the gauge variables to the underlying physical fields may range from the direct -- for example in the case of superconductivity, where gauge fields emerge to explicitly reveal the independence of physical observables upon the phase of the wave function -- to the indirect -- for example in the case of slave fields used in strongly correlated electron systems, where the electrons are split or fractionalised and gauge fields emerge as effectively Lagrange multipliers imposing the constraint that the system is in fact made of electrons. In magnetically or orbitally ordered materials, gauge fields prove a convenient way of characterising the Berry phases acquired by an electron propagating through the system or of the emergence of topological defects in the magnetic order near to a critical point.

A complete discussion of this fascinating diversity of ideas is far beyond the scope of this article. The salient point, however, is that the challenge of trying to understand more and more complicated behaviour has led to the development of tools that are ever more like those being developed at the cutting edge of high energy field theory.
Of course, progress in condensed matter and high energy physics have always gone hand in hand. Ideas like spontaneous symmetry breaking and the Higgs mechanism, and the last great meta-theory of physics, the renormalisation group, were developed more or less in parallel. What is a surprise, however, is just how far the gauge-gravity duality seems to take us from our starting point. It involves not just at a different mathematical model, but a different {\it type} of setting entirely. This mapping is usually formulated using mathematical  tools that are unfamiliar to the majority of condensed matter physicists. Nevertheless, it follows from some fundamental and well-known principles.

\subsection{Gauge-String Duality in Classical and Quantum Physics}

The gauge-gravity duality is founded upon a familiar feature of classical electromagnetism - the duality between a description in terms of field strengths and lines of force or flux. These dual perspectives are so thoroughly embedded in our understanding  that we hardly give them a second thought\cite{Maxwell}. However, it is precisely the quantum mechanical version of this duality that underpins the mapping from condensed matter systems to gravitational systems\cite{Gauntlett:1998}.

\begin{figure}[ht]
\hspace{-0.9 \linewidth}
a)
\\
\includegraphics[width=1.\linewidth]{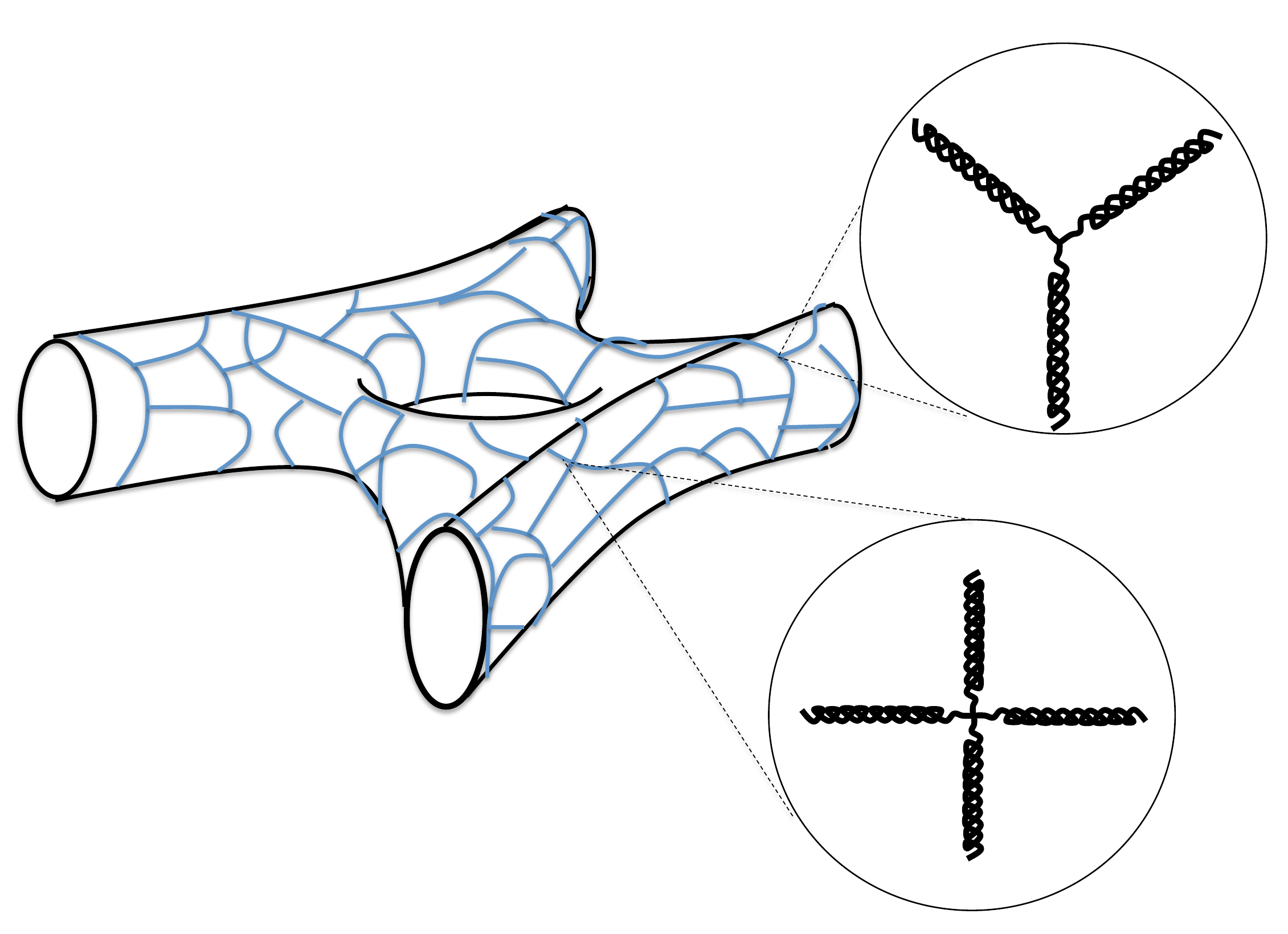}
\\
\hspace{-0.9 \linewidth}
b)
\\
\includegraphics[width=1.\linewidth]{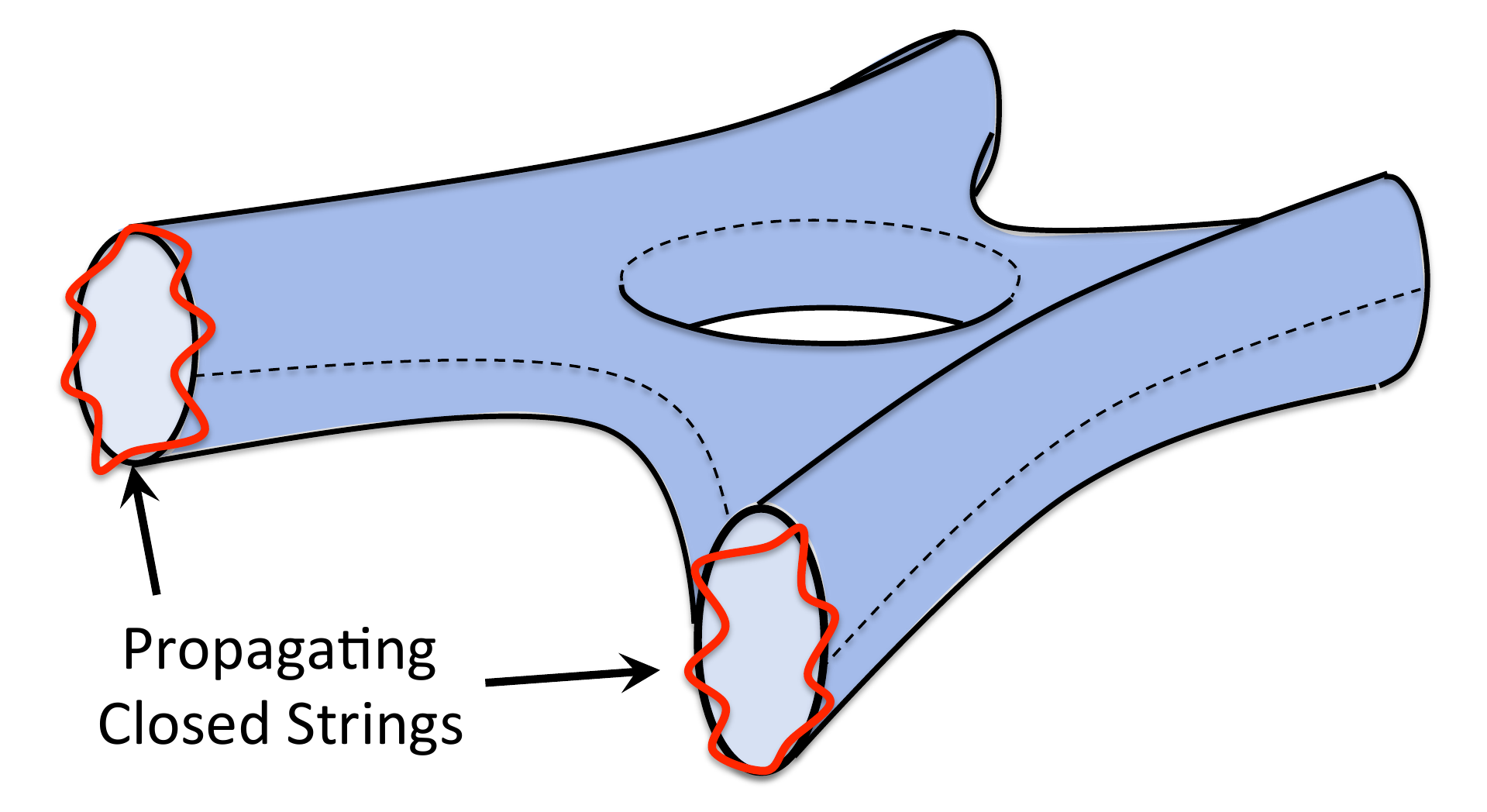}
\caption{{\bf Feynman Diagrams for QCD map to those of Closed Strings:} a) The planar Feynman diagrams of QCD constructed from the gauge interactions can be classified in terms of the surfaces that they triangulate. b) This classification puts them into one-to-one correspondence with the Feynman graphs for the propagation of closed strings. Here, the interaction vertices for QCD are shown together with a Feynman diagram for the interaction of two closed strings.}
\label{fig:QCDvsStrings}
\end{figure}

If we view electromagnetism in terms of field strengths, the fundamental constituents are the point-like value of fields at a given point in space. On the other hand, flux lines are one-dimensional, string-like objects. The possibility of interchanging these two viewpoints on a classical level convinced theoretical physicists such as t'Hooft, and Polyakov 
that there should be a similar duality between quantum gauge fields and strings --- a quantum theory of flux lines is in essence a string theory\cite{THooft,Polyakov1978477,PolyakovBook,GrossI,GrossII}\footnote{As an amusing historical aside, the notion that topologically non-trivial states of gauge fields - and contemporary work on knot theory by Tait\cite{Tait} - led Lord Kelvin to attempt to formulate a theory of fundamental particles as knots in the aether\cite{KelvinKnots}. }. There were many early hints that this expectation was correct. For example, Feynman diagrams describing the way in which gauge fields interact can be classified in terms of the genus of the surface that they triangulate [see \figref{fig:QCDvsStrings}]. In turn, the equivalent of Feynman diagrams to describe the interaction of strings form surfaces of different genus. The sets of surfaces derived in these two constructions are in one-to-one correspondence, strongly suggesting that the the underlying gauge and string theories are also in correspondence\cite{THooft,Polyakov1978477,PolyakovBook,GrossI,GrossII}.

This correspondence, though tantalisingly close, remained elusive for several decades. The final details required to make it mathematically rigorous could not be found. The insight that  allows all of this to fall into place came in 1997 when Maldacena was able to demonstrate a correspondence between operators in a particular supersymmetric gauge theory and a string theory\cite{MaldacenaI}. Crucially, in order for all of this to work, the string theory lives in higher dimension than the gauge theory. Very rapidly, the notion of correspondence between gauge and string theories crystallised around this idea\cite{GubserKlebanovPolyakov,Witten}.

\vspace{0.1in}
{\it Holography and Reduction to Gravity:}
An outstanding feature of string theory is the very natural way in which gravity emerges from it -- almost as if compelled by its very mathematical elegance rather than by design. The classical configurations of string theories are Einsteinian gravitational theories.
This fact is crucial to the utility of gauge-gravity duality for calculations. As it stands, we have motivated a mapping between a quantum gauge theory and a quantum string theory. If we had to calculate the full quantum behaviour of the string theory, there would not be much advantage in the mapping. However, for string theories whose behaviour is dominated by their classical, gravitational configurations the task is considerably simpler (we will comment upon a class of theories for which this is the case shortly). Calculations are reduced to that of evaluating propagators in the background of this classical curved space time. 

It has become conventional amongst string theorists to refer to this mapping between a $d$-dimensional field theory and a higher (usually) $d+1$-dimensional gravitational dual as holography. In the same way as a hologram contains all of the information about the shape of a $3$-dimensional object in $2$-dimensions, the gauge/field theory contains all of the information of the string theory in fewer dimensions. This is the result of some very deep principles whose consequences are still being unpacked. The implications for condensed matter are exciting. 

Indeed, the idea that consistent quantum theories of gravity should be holographic  has been understood for some time. The maximum information in a region of space is proportional to its surface area. This is a constraint forced upon us by thermodynamics and is precisely the line of thinking that led Bekenstein and Hawking\cite{Bekenstein:1973zg,HawkingI,HawkingII} to deduce the existence of black hole entropy and radiation.
Black holes play a crucial role in the holographic mapping also. A field theory at temperature, $T$, maps to a gravitational bulk that contains a black hole characterised by the same temperature. 

\begin{figure}[ht]
\includegraphics[width=1.\linewidth]{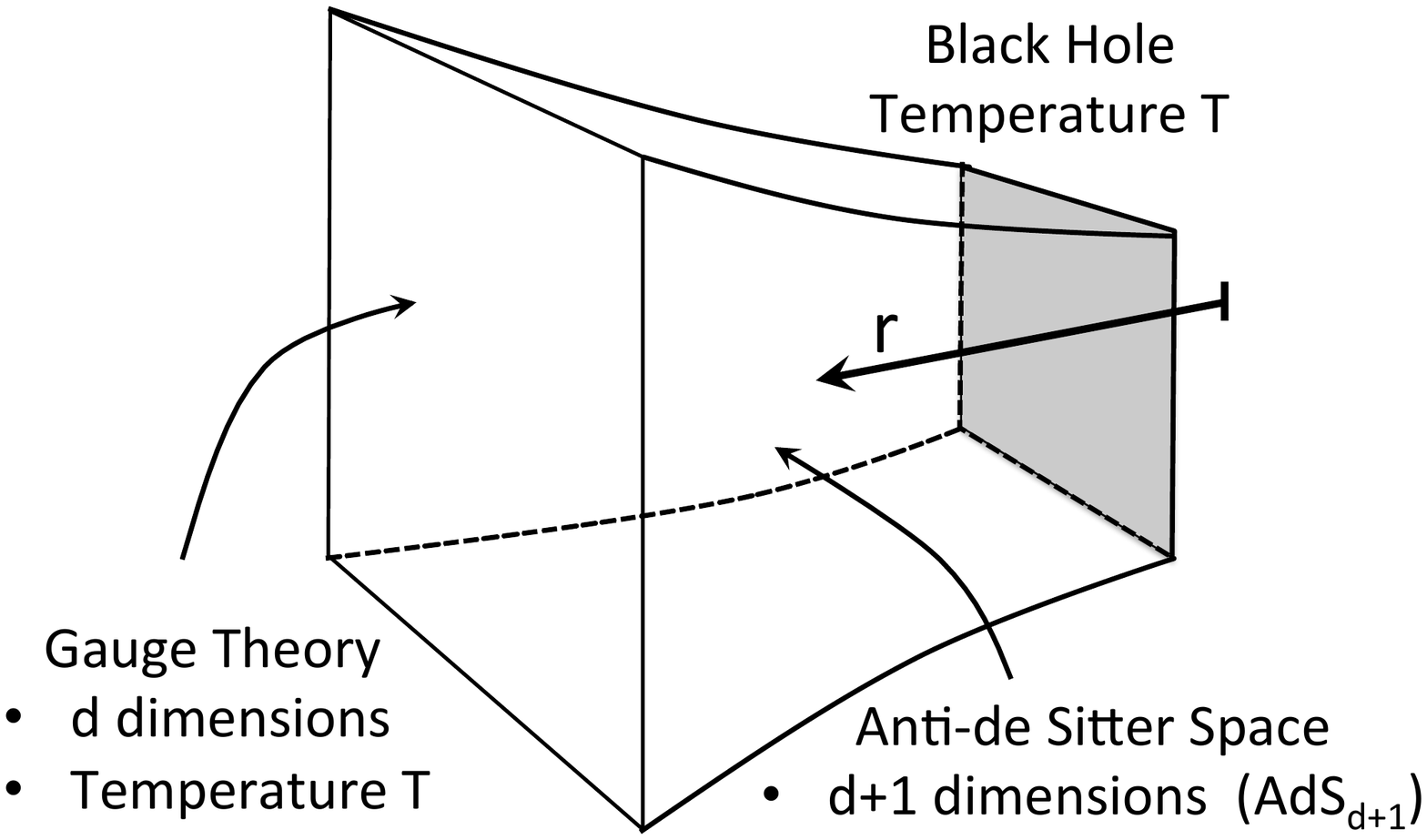}
\caption{{\bf Cartoon Representation of Gauge-Gravity Duality:} Eq.(1) displays an equivalence between a gauge theory in d dimensions and a string theory in d+1 dimensions. The classical saddle point of the latter is an Anti-de Sitter space with horizon characterised by the same temperature are the boundary theory. The negatively curved Anti-de Sitter space is often indicated as shown.}
\label{fig:GaugeGravityDuality}
\end{figure}

\vspace{0.1in}
{\it Calculating with gauge-gravity duality:}
So how does all of this fit together into a calculation? The aim of this review is not to show in detail how to perform calculations -- for that, the reader is directed towards one of the many reviews \cite{HartnollReview,SachdevReview}.
Nevertheless, in order to make the subsequent discussion more concrete, a little more detail is in order.  The  mathematical relationship that  forms the basis for calculations using the gauge-gravity duality  is a mapping between the generating function of the gauge theory  and the partition function of a string theory\cite{MaldacenaI,ADSCMTBook,HartnollReview,SachdevReview}. 
In limit where the mapping is useful, the string theory is dominated by its classical gravity configurations.

The generating function of the field theory is just the partition function - the sum of the Boltzmann weights for all possible field configurations - calculated in the presence of some external source field. Consider, for example,  that we are interested in the behaviour of currents ${\bf j}({\bf x},t)$.
The generating function will be a functional of source fields ${\bf A}({\bf x},t)$ that couple to the currents and is given schematically by
\begin{equation}
Z[{\bf A}({\bf x},t)]= Tr \; e^{-\beta\left[ {\cal H} - \int d{\bf x} {\bf A}.{\bf j} \right]},
\label{GeneratingFunction}
\end{equation}
where $\beta$ is the inverse temperature. The trace is taken over the degrees of freedom of the field theory, whose dynamics are determined by the Hamiltonian ${\cal H}$. The current ${\bf j}({\bf x},t)$ is a function of these same degrees of freedom. Eq.(\ref{GeneratingFunction}) is referred to as the generating function because it may be used to calculate (or generate) correlation functions of currents by taking its derivatives with respect to 
 ${\bf A}({\bf x},t)$.

The partition function of a string theory is similarly given by the sum of Boltzmann weights for all possible configurations of strings and fields. 
We wish to consider a limit where the string theory is dominated by its classical gravity configurations.The generating function of the field
theory and the partition function of the gravity dual, $ {\cal Z}[ {\cal A}({\bf x},r,t)]$ are related to one another according to 
\begin{equation}
Z[{\bf A}({\bf x},t)]
=\left. {\cal Z}[ {\cal A}({\bf x},r,t)]\right|_{{\cal A}({\bf x},r=\infty, t)={\bf A}({\bf x},t)}
\sim
e^{-{\cal S}_{cl}}.
\label{Duality}
\end{equation}
The fields ${\cal A}({\bf x},r,t)$ are dynamical gauge fields in the dual gravity theory whose values on the boundary are equal to the source fields of the generating function for the boundary field theory. The coordinate $r$ measures the distance along the extra dimension into the bulk gravitational theory [see Fig.\ref{fig:GaugeGravityDuality}]. 
As indicated in Eq.(\ref{Duality}), if the string theory is dominated by its classical gravity configurations its partition function reduces to the exponential of the classical action of these configurations, ${\cal S}_{cl}$. The evaluation of the right hand side is reduced to the solution of classical equations for the metric and bulk gauge fields. As a result, scaling properties emerging from complex quantum many-body correlations in the boundary field theory have been mapped to the general covariance of the dual gravitational metric. 

A calculation using gauge-gravity duality, then, typically involves solving classical equations of motion for the fields ${\cal A}({\bf x},r,t)$ in a metric that reflects the symmetries of the boundary theory. 
The existence of such a mapping places particular demands upon both the nature of the gauge theory and of the gravitational metric. At the very least, unlike the space that we inhabit, the metric must have a boundary upon which the gauge theory lives.  Metrics with a negative curvature have such a boundary at spatial infinity. The simplest example is anti-de Sitter space  (AdS -- the source of   one of the common banners under which this field has marched). The classical theory combining gravity and electromagnetism\footnote{The action for the Einstein-Maxwell theory is given by
$
{\cal S}
=
\int d^{d+1} x \sqrt{-g}
\left[
\frac{1}{16 \pi G}
\left(
R+ \frac{d(d-1)}{L^2}
\right)
-
\frac{1}{4 e^2} F^2
\right].
$
$G$ is Newton's gravitational constant and $L$ is the (negative) radius of curvature of the resulting space. $R$ is the Riemann curvature scalar and $g$ the determinant of the metric. $F$ is the flux associated with the gauge fields ${\cal A}$. }
 with a negative cosmological constant, has a solution with ${\cal A}({\bf x},r,t)=0$ and a background metric given by
\begin{eqnarray}
ds^2
&=&
 \frac{L^2}{r^2}
 \frac{dr^2}{f(r)}
 +
 \frac{r^2}{L^2} \left(
 -f(r)dt^2 +d{\bf x}^2
 \right)
 \nonumber\\
f(r)&=&
\left(
1-
\frac{r_h}{r} \right)^d,
\label{Metric}
\end{eqnarray}
where $L$ is the (negative) radius of curvature of the resulting space-time and $r_h$ is the location of a black hole horizon. The black hole -- which is in this case a planar black hole as can be seen from the form of the metric in ${\bf x}$ -- has a Hawking temperature $T= 3 \hbar c r_h/(4 \pi L^2)$ associated with it [see \figref{fig:GaugeGravityDuality}]. Crucially, the boundary theory has the same temperature. 

In order to make use of the mapping Eq.(\ref{Duality}), we must solve the  the Einstein-Maxwell equations with boundary conditions ${\cal A}({\bf x},r=\infty, t)={\bf A}({\bf x},t)$. In general we expect that the resulting non-zero value of 
${\cal A}({\bf x},r, t)$ would also modify the metric. Under the assumption that this effect is negligible, solving the Maxwell equations \footnote{With the important addition of leading fourth order gradient terms to the action
$$
{\cal S}_{\hbox{4th order}}
=
\int d^{d+1} \sqrt{-g} \frac{\gamma L^2}{e^2} C_{abcd} F^{ab} F^{cd},
$$
where $\gamma$ is a parameter and $C_{abcd}$ is the Weyl curvature tensor, 
a phenomenology encompassing a broader class of problem, including the Bose Hubbard model can be studied. See \cite{Myers:2011po}.}
 for ${\cal A}$ in the background metric given by Eq.(\ref{Metric})  allows current correlations in the boundary field theory to be calculated using Eq.(\ref{Duality}). More complicated metrics -- for example, in which the black hole is charged or where a richer mix of fields are allowed to propagate in the bulk -- can be used to model different boundary theories. The interplay of additional fields in the bulk with the metric can lead to a breakdown of Lorentz invariance of the gravity dual, reflecting different types of critical scaling in the boundary theory.

\vspace{0.1in}
{\it Symmetries and Scaling:}
At this point, it is worth stepping back to make a few comments. The string and gauge theories for which Maldecena originally conjectured his mapping were ones with additional symmetry between fermions and bosons and, importantly, a large number of colours ($N$ say) of gauge fluctuations. The latter acts to stabilise the classical gravitational configuration --- in essence, a macroscopic occupation of modes gives rise to a classical limit in the usual way. The consequences of this supersymmetry and large-N have yet to be fully disentangled\footnote{One may take the view that - though of different mathematical origin -  large $N$ plays a conceptually similar role to the (vector) large $N$ in non-perturbative field theoretical treatments of strongly correlated systems, such as the $O(N)$ extension of the $O(2)$-model, for example.}. We shall return to this when we discuss out-of-equilibrium aspects of the theory. In the meantime, suffice it to say that there is much circumstantial evidence that the mapping between gauge theory and gravity has validity that extends beyond the particular string theory in which it is embedded. We shall assume that this is the case and discuss its consequences. In this view, the enhanced symmetry of the original model simply serves to make the calculations demonstrating the mapping tractable.

Nevertheless, some of the symmetries are certainly important for the mapping to work. The gauge theories for which this mapping works are of a particular type, with spatial and temporal scaling of correlations, corresponding to those found in quantum critical condensed matter systems\cite{Sachdev1999,RevModPhys.69.315}. 
Quantum criticality occurs when a system is tuned to a zero-temperature phase transition as a function of some control parameter such as external magnetic field or pressure. Loosely, the transition is driven by a competition between quantum fluctuations and interactions, rather than between thermal fluctuations and interactions as at a classical phase transition. Near to the transition point, correlations show power-law dependences upon separation in space and time. These scaling laws are the emergent result of complicated many-body correlations in the gauge theory. The gauge-gravity duality maps them to the fundamental invariance of the gravitational metric -- general covariance. 
With this mapping in hand, calculational tools and physical intuition developed for classical gravity may be used to reveal unexpected properties of the quantum critical system.

Examining the metric can give us some insight about why this construction works\cite{SachdevReview}. In the limit of zero temperature, $r_h \rightarrow 0$ and the function $f(r)=1$. Eq.(\ref{Metric}) then  corresponds to anti de Sitter space. As noted above, this space possesses a boundary upon which we may view the gauge theory as living. A useful and frequently used cartoon of anti-de Sitter space is provided by Circle Limit I and II of M. C. Escher (not included here for copyright reasons). The anti-de Sitter metric is invariant under scale transformations

\begin{eqnarray*}
t & \rightarrow& t/b, \\
{\bf x} & \rightarrow & {\bf x}/b, \\
r & \rightarrow & r b.
\end{eqnarray*}
This scaling has two features of particular note. Firstly, $t$ and ${\bf x}$ scale in the same way. The dynamical exponent, $z$, counts the number of effective spatial dimensions that time contributes to the theory so that in general $t^{1/z}$ scales in the same way as ${\bf x}$. In the present case $z=1$. Recent developments have attempted to extend the gauge-gravity duality to theories with dynamical exponents not equal to one. Secondly - and more importantly for our general interpretation of gauge-gravity duality - 
notice that the scaling of the new coordinate $r$ is different from that of space and time in the boundary theory. It scales in the same way as $1/t$ or $1/{\bf x}$, {\it i.e. } in the same way as energy or momentum. Different depths of $r$ correspond to different energy scales with the low-energy/ infra-red near $r=0$ and the high-energy/ultra-violet near the boundary at $r=\infty$. The gauge-gravity duality has transformed the renormalization invariance of the quantum critical theory into a geometrical invariance of  the metric. We will return to this idea later. It forms the basis of attempts to construct the mapping from a condensed matter system  to a gravity dual directly without an explicit embedding in a string theory.

\section{Applications of Gauge Gravity Duality}

The ideas described above may seem rather exotic and it is tempting for a conservative  physicist to be sceptical about their usefulness in condensed matter. For now, let us suspend our disbelief and see where Eq.(\ref{Duality}) and the gauge-gravity duality might take us. In this section, I will attempt to give a flavour of what has been achieved so far. I will start with a discussion of the application of gauge-gravity duality in high energy physics, before going on to discuss the application to condensed matter. 
I will divide the latter into four parts: bosonic systems, fermonic systems and their instabilities, entanglement and the behaviour of quantum critical systems out of equilibrium. 

\subsection{High Energy Physics}

Heavy ion collisions and the resulting hot quark-gluon plasma allowed string theory to make the transition from theoretical abstraction to experimental realisation. These experiments put QCD in a regime where perturbative calculations are futile. In particular, the observed viscosity of the quark-gluon plasma was much lower than had been anticipated on the basis of perturbative calculations\cite{QCDViscocity}.

That the essence of this seemingly counterintuitive result is now clear is in large part down to the contribution of holographic methods\cite{Kovtun:2005xg}. A low viscosity is not that surprising - strong scattering naturally leads to low viscosity as it rapidly transfers momentum between layers of a sheared fluid. The contribution of holography was to show that there is a universal scale for this viscosity that is almost saturated by the quark-gluon plasma - it is an almost perfect fluid.

The holographic calculation does not treat QCD directly. Rather, it treats the behaviour of a particular supersymmetric extension of QCD. However, the insights that it provides about the universal limit upon fluid viscosity are quite general. Indeed, the relativistic hydrodynamic equations that result include terms that were previously missing from the canon on hydrodynamic flow. The result is a bound upon the ratio between the shear viscosity, $\eta$, and the entropy density, $s$,
\begin{equation}
\frac{\eta}{s} = \frac{\hbar}{4 \pi k_B}.
\label{viscocityLimit}
\end{equation}
The central message that the condensed matter physicist can take from these results is that a holographic analysis can provide quantitative bounds that can transcend the particular realisation in a supersymmetric gauge theory. With hindsight, of course, it is often possible to get at these results without the change of perspective afforded by the holographic approach. The limit of viscosity can be thought of as a sort of Mott-Ioffe-Regel\cite{IoffeRegel,MottI,MottII} limit where the maximum scattering rate is attained for particles with the thermal de Broglie wavelength. In a quantum critical fluid, both the thermal wavelength and scattering rate become universal functions with temperature providing the only energy scale. For certain responses, this energy scale drops out and one is left with a universal viscosity. The nice feature of the holographic analysis is that it shows how these types of result may be deduced from emergent scaling rather than microscopic considerations. 

\subsection{Condensed Matter}

\subsubsection{Bosons}

{\it The Bose-Hubbard model} is one of the  benchmarks for the study of quantum criticality. It describes bosons, hopping around a lattice with an on-site repulsive interaction. As the ratio between hopping and repulsive interaction is tuned, the model shows a transition from an insulator to a superfluid. The study of the quantum critical point between the two has lead to the development of much of the toolbox of bosonic quantum criticality\cite{Sachdev1999}.

The development of our understanding of this model has a long history. Early works classified the behaviour of this system in terms of types of dynamics that the model displayed\cite{Hohenberg:1977yy}, focussing upon cases where this dynamics was over damped. The developments feeding most directly to the subject of this review follow Refs.\cite{Fisher:1989uj,Cha:1991cz} and others concerning the behaviour of a system with intrinsic relativistic dynamics. 

These attempts to determine the response functions of this system involved positing a field-vortex duality in order to determine the critical conductivity. Damle and Sachdev\cite{PhysRevB.56.8714} pointed out that caution is required as the limits of temperature and frequency going to zero do not commute for such systems. Also, although the relativistic Bose-Hubbard model in d spatial dimensions may be mapped to a model of classical dynamics in d+1 dimensions, the rotation to imaginary time involved in this mapping does not commute with any of the approximation schemes used in calculations. Damle and Sachdev provided an alternative scheme that invokes a controlled expansion in the limit where probe frequencies are much less than intrinsic frequencies of the system - the hydrodynamic limit.

The essence of the Damle Sachdev approach is to develop a Boltzmann transport description of the distribution functions of 
normal modes, controlling the critical scattering between them by means of an $\epsilon$ or $1/N$ expansion. The Boltzmann equation determined in this way is deceptively simple. 
All of the effects of quantum criticality are contained in the linearised scattering integral ${\cal F}_{{\bf k},{\bf q}}$ and the dispersion of the normal modes. Explicit forms for ${\cal F}_{{\bf k},{\bf q}}$ were given by Damle and Sachdev within the $\epsilon$-expansion\cite{PhysRevB.56.8714} (and later the $1/N$ expansion). A number of interesting properties of and relationships between response functions are contained within the detailed form of this scattering integral. 

{\it Relationships Between Response Functions:} Though carried out with other aims, this work opened the door to comparisons with the holographic approach. As noted above, the holographic approach had proven adept at treating relativistic hydrodynamics and revealed some unexpected features of the behaviour of the quark gluon plasma. This same approach -- essentially studying the Einstein-Maxwell system introduced at the end of section IA - has been applied with some success to condensed matter systems\cite{Herzog:2007mu} in general and to graphene in particular \cite{Muller:2009wa}, which was found to be a perfect fluid in the same manner as the quark-gluon plasma. In Ref.\cite{Hartnoll:2007yq} Hartnoll {\it et al.}  studied the hydrodynamic limit of this model in the presence of a magnetic field, with the aim of capturing phenomenology corresponding to that of the Bose-Hubbard model in crossed electric and magnetic fields. The constraints of relativistic invariance and the hydrodynamic limit allowed them to deduce a startling relationship between the thermal conductivity, $\kappa$, electrical conductivity, $\sigma$, entropy density, ${\cal S}$, magnetic field, $B$ and temperature $T$;
\begin{equation}
\kappa = \frac{T {\cal S}^2}{B^2 \sigma}.
\label{KappaDuality}
\end{equation}
In fact, a close precursor to this result was found  using the Boltzmann approach\cite{PhysRevLett.98.166801} and indeed Eq.(\ref{KappaDuality}) is recovered exactly in the hydrodynamic limit\cite{Bhaseen:2009ec}. In the latter, this result can be traced back to a near zero mode of the scattering integral corresponding to energy conservation in the scattering process. This, coupled to the scaling of the scattering integral with temperature, leads to the above relationship. 
The holographic approach on the other hand deduces this result as a consequence of the general covariance of the gravitational dual - {\it i.e.} ultimately, the emergent scale invariance of the quantum critical system. 

{\it Sum Rules:} Investigations in this direction continue to reveal some curious and poorly understood properties of bosonic quantum criticality, and have extended the validity of dualities like that contained in Eq.(\ref{KappaDuality}). A very recent preprint of Witczak-Krempa and Sachdev\cite{Witczak-Krempa:2012gi} has demonstrated the existence of frequency sum rules obeyed by the quantum critical conductivity. These occur as the result of quasi-normal modes that dominate the dynamics of the gravity dual. The interpretation of these results within the Boltzmann approach is not yet fully understood. Witczak-Krempa and Sachdev re-analyse the Boltzmann transport equations within a $1/N$ expansion and find that the frequency-dependent form reduces to a very simple form --- a form that these authors refer to as quasi exact. It seems likely that understanding the physics underpinning these results will provide further revealing insights about the nature of quantum critical transport.

{\it Supersymmetry and Bosonic Response:} The results discussed so far have been obtained from the gravity dual by calculating classical bosonic propagators within the bulk of the dual metric. The main question that this prompts concerns the embedding of this within a theory with extended supersymmetry. Repeating the caveat that we have made several times now, the holographic calculation is not for the Bose-Hubbard model directly. However, the spirit of the approach is to aim to reveal physics whose applicability extends beyond the particular embedding. The Bose-Hubbard model plays a very important role in this game --- because it is amongst those quantum critical theories over which we have most analytical control, it acts as a good benchmark. To date, all of the results that have been compared are in agreement, both in and out of equilibrium.  With this assurance in mind, we will now turn to look at some of the more challenging and speculative applications of the gauge-gravity duality.

\subsubsection{Fermions and Superfluids}

The mission to understand the general organising principles governing collective quantum behaviour has proven a recurrent challenge for condensed matter physics\cite{AndersonBasicNotions}. Facing it has led to some of physics' most profound insights, ideas such as Landau-Fermi liquid theory, spontaneous symmetry breaking and the renormalisation group. Together these ideas, and the mathematical techniques to implement them, allow the description of a very broad range of phenomena. 
However, a large and increasing number of metals and magnets have been discovered whose behaviour cannot be explained using these conventional tools. 

New ideas and tools are clearly required and holography provides an intriguing and welcome possibility. This is a delicate task. The very nature of strongly correlated metals  is such that we do not have a complete understanding or analytical control of their behaviour using the current tools of condensed matter physics. Appropriate benchmarks are largely absent and much care is required in the interpretation of results. Nevertheless, a lot of progress has been made and the field is moving forward rapidly. There are encouraging reasons to believe that the approach can indeed lead to new insights and alternative descriptions of unusual metallic and superfluid phases.

{\it First Attempts - The Marginal Fermi Liquid:}
The first attempts to calculate the behaviour of fermonic systems from the holographic point of view proceeded along very similar lines to the bosonic calculations outlined above\cite{Lee:2009tr,Liu:2011vf,Faulkner:2011oh,Cubrovic}. Firstly, a choice of  background metric must be made that reflects the symmetries anticipated for the strongly correlated fermonic system. An appropriate choice in this case is a solution of the Einstein-Maxwell gravity theory containing a black hole with both mass and charge. The charge leads to a finite electrical flux threading the boundary. ${\cal A}({\bf x},r=\infty,t)= \mu$ and therefore implies a finite charge density in the boundary theory. 

The background metric obtained in this way is the Reissner-Nortstr\"om metric which takes the same form as that given in Eq.(\ref{Metric}) with 
$$
f(r)= 1+ \lambda (r_h/r)^4 - (1 + \lambda) (r_h/r)^3,
$$
where $\lambda= 4 \pi L^2 G \mu^2/(e^2 r_h^2)$ and $G$ is Newton's constant. This metric also supports a non-zero scalar potential ${\cal A}({\bf x},r,t)= \mu(1-r_h/r)$. A classical Dirac equation is solved in this background space-time in order to determine fermonic correlation functions of the boundary theory. This was met with some early success, making connection to a particularly intriguing, strange metallic phenomenology. 

\begin{figure}[ht]
\includegraphics[width=1.\linewidth]{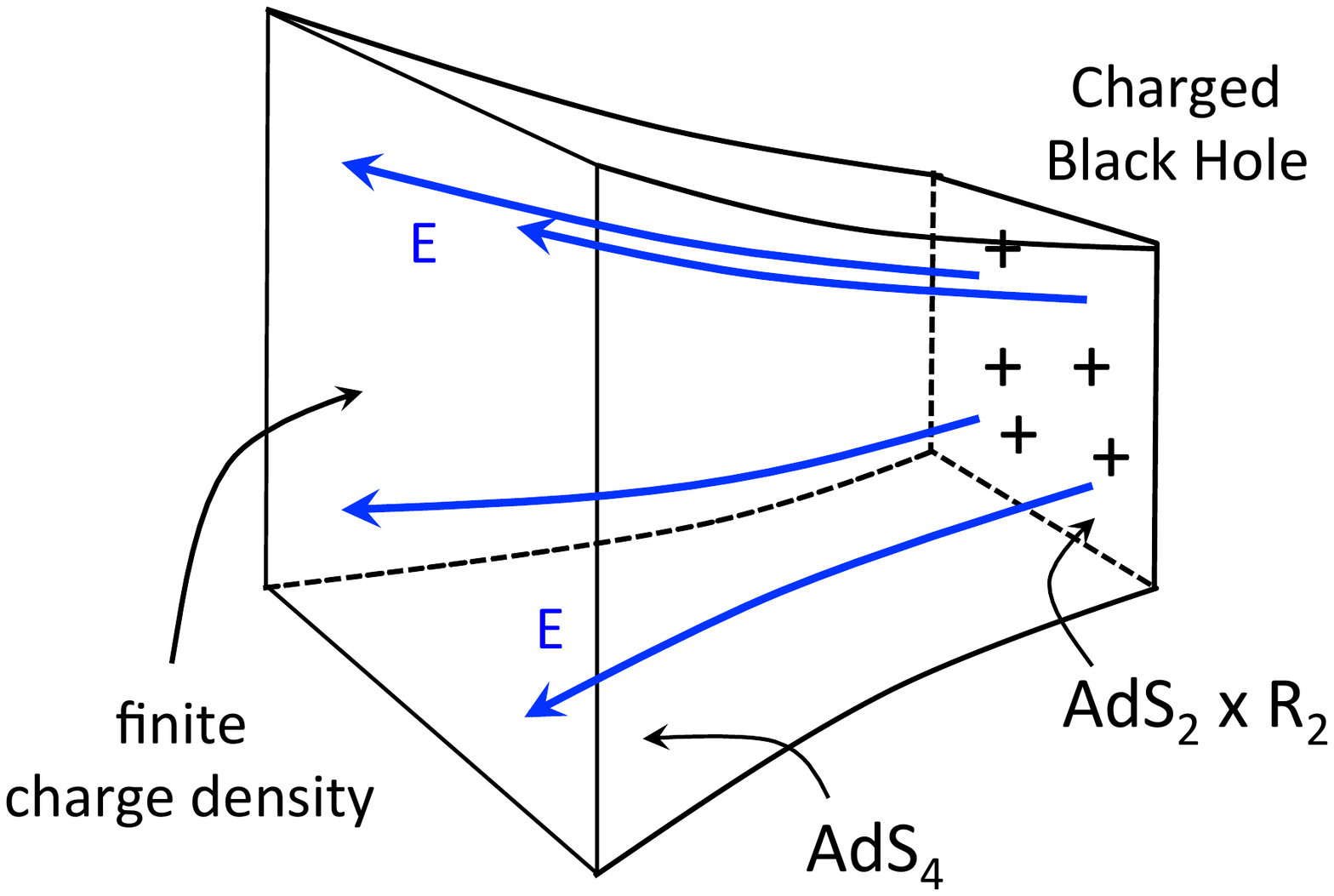}
\caption{{\bf The Holographic Strange Metal:} Allowing a charged black hole leads to an electric flux through the boundary. This induces a finite fermion density in the boundary. It also changes the geometry of the metric. Near the horizon, the space changes from $AdS_4$ to $AdS_2 \times R_2$ encoding critical scaling in frequency alone and so marginal Fermi liquid behaviour.}
\label{fig:HolographicStrangeMetal}
\end{figure}

Near the boundary at $r\rightarrow \infty$, the metric with this modification takes the same asymptotically $AdS_4$ form as before; $ds^2=(L^2/r^2) dr^2 +(r^2/L^2)(d{\bf x}^2 -dt^2)$. However,  it is markedly different near to the horizon, taking the form
$$ds^2=(L^2/r_h^2) dz^2/(6 z^2)-6(r_h^2/L^2) z^2 dt^2+ (r_h^2/L^2) d{\bf x}^2,$$
where $x=r-r_h$. The space factorizes into a two-dimensional anti de Sitter ($AdS_2$) form in  $z,t$-space and a flat ($R_2$) form in the spatial coordinates of the boundary [See \figref{fig:HolographicStrangeMetal}]. This different behaviour for space and time has profound consequences for fermonic propagation in the boundary theory. Critical scaling is found in the frequency dependence alone and in certain parameter ranges can lead to marginal Fermi liquid behaviour of the boundary propagators, with a self-energy given by
\begin{equation}
\Sigma({\bf k},\omega)  \sim \omega \ln \omega.
\label{MarginalSelfEnergy}
\end{equation}
The marginal Fermi liquid was initially postulated as a phenomenological way to capture a broad range of experimental signatures found in cuprate superconductors\cite{Varma:1989ln}. Attempts to derive this phenomenology from microscopic Hamiltonians have proven difficult with no consensus yet having be attained. Finding similar phenomenology in the holographic approach generated great initial excitement.

{\it Diverging Entropy and Fermi-Liquid Instabilities:} Fairly rapidly, it became apparent that this approach suffers from some the same problems as attempts to recover marginal Fermi liquid behaviour from condensed matter Hamiltonians. A central feature of the marginal Fermi liquid is that the self-energy depends only upon frequency\cite{Varma:1989ln}. This is the main assumption of dynamical mean-field theory\cite{Georges:1996cz}. This approach has been applied to a variety of problems such as the Mott transition and of, particular note here, to the prediction of so called local quantum criticality\cite{LocalQC} --- quantum criticality where the scaling is in time alone. 

The essence of calculations using dynamical mean field theory is to determine the behaviour at a point in the system self-consistently by determining the effect of its interaction with a dynamical mean field bath made with identical local behaviour. This effectively says that every point in the system is coupled to its own bath and massively over-counts the degrees of freedom. The dynamical mean field treatment, therefore has a macroscopic zero-temperature entropy.

The generalised marginal Fermi liquid of the holographic approach also has such a macroscopic, zero-temperature entropy. This shows up in a curious way. The mapping between the quantum critical boundary and the gravity bulk identifies the temperature of the boundary theory with the temperature of the black hole. For the Reissner-Nordstr\"om black hole, this is given by 
$k_B T = (3 \hbar c r_h/4 \pi c^2)(1-\lambda /3)$ with the striking result that the horizon remains at finite radius ($\lambda \sim 1/r_h$) when $T\rightarrow 0$. Since, following Bekenstein-Hawking, we know that the black hole has an entropy proportional to its area, we deduce a finite entropy density as $T\rightarrow 0$. 
There is an appealing parallel between this and the dynamical mean field ansatz.  The AdS$_2$ metric may be thought of as a encoding local quantum criticality by imposing scaling only in the temporal direction. The gravitational embedding of this suggests that the propagators of a locally quantum critical theory are forced by consistency to take a generalised marginal form.  

\begin{figure}[ht]
\includegraphics[width=1.\linewidth]{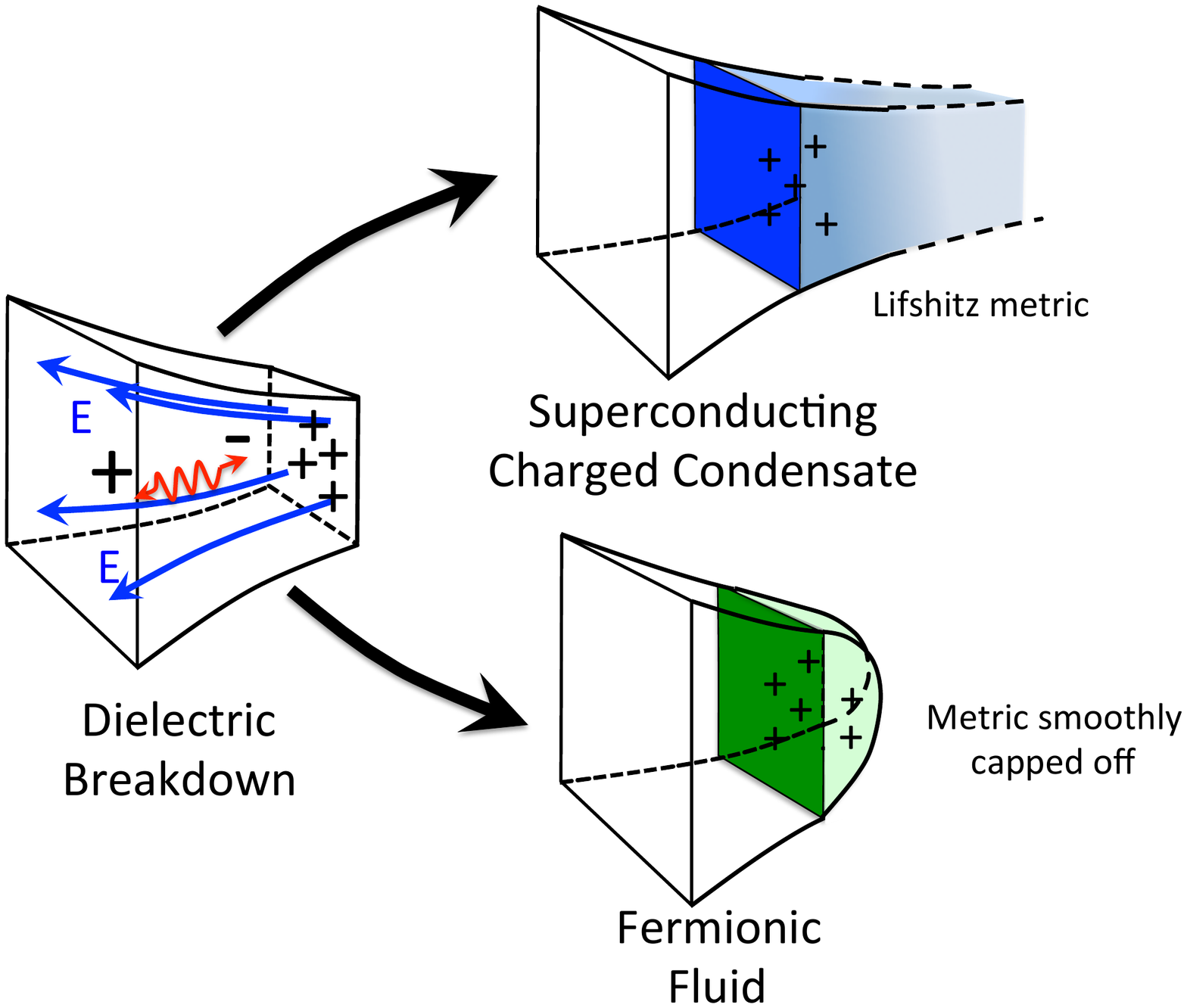}
\caption{{\bf The Electron Star and the Holographic Superconductor:} The electric field at the horizon of the charged black hole can become sufficiently large that dielectric breakdown of the vacuum occurs outside of the horizon. Pairs of positive and negative charges are produced. The positive charges are attracted to the negatively charged black hole and the negative charges repelled from it. This screening of the black hole may occur by the formation of a charged condensate in the bulk - corresponding to a superconductor in the boundary theory - or by the formation of a charged fermonic fluid - corresponding to the formation of a non-Fermi liquid in the boundary theory.  The back reaction of these fluids may modify the metric in several ways, a couple of examples of which are illustrated.}
\label{fig:ElectronStarandHolographicSuperconductor}
\end{figure}

The way in which the macroscopic degeneracy of the zero-temperature horizon is negotiated also takes inspiration from condensed matter. It turns out that that if one studies the gravitational metric carefully, as the temperature of the black hole is reduced, below some scale the metric corresponding to the marginal Fermi liquid is no longer the minimal solution. Below this temperature scale a new metric takes over with finite expectation of a charged scalar field in the bulk corresponding to a superconducting order parameter in the boundary theory [see \figref{fig:ElectronStarandHolographicSuperconductor}]. These results have been confirmed both in a top down\cite{Gauntlett:2009yy,GubserTopDown} (working within a known string theoretical embedding) and bottom up\cite{Hartnoll:2008sp} (working with a metric with appropriate symmetry and assuming that there is a consistent - if unknown - string theoretical embedding) setting. The instability of the marginal Fermi liquid to superconductivity then seems to be on a firm footing. 

The metric that results after the formation of the holographic superconductor has a rather different form to those discussed up to now. It no longer posses a horizon and has  different scaling in space and time. The formation of this metric -- which has become known as the Lifshitz metric --  is illustrated schematically in \figref{fig:ElectronStarandHolographicSuperconductor}. The properties of the critical superconductor that results are different from other types of superconducting phases that can potentially be observable with spectroscopic probes\cite{ZaanenSpectroscopy}.

\begin{figure}[ht]
\includegraphics[width=0.8 \linewidth]{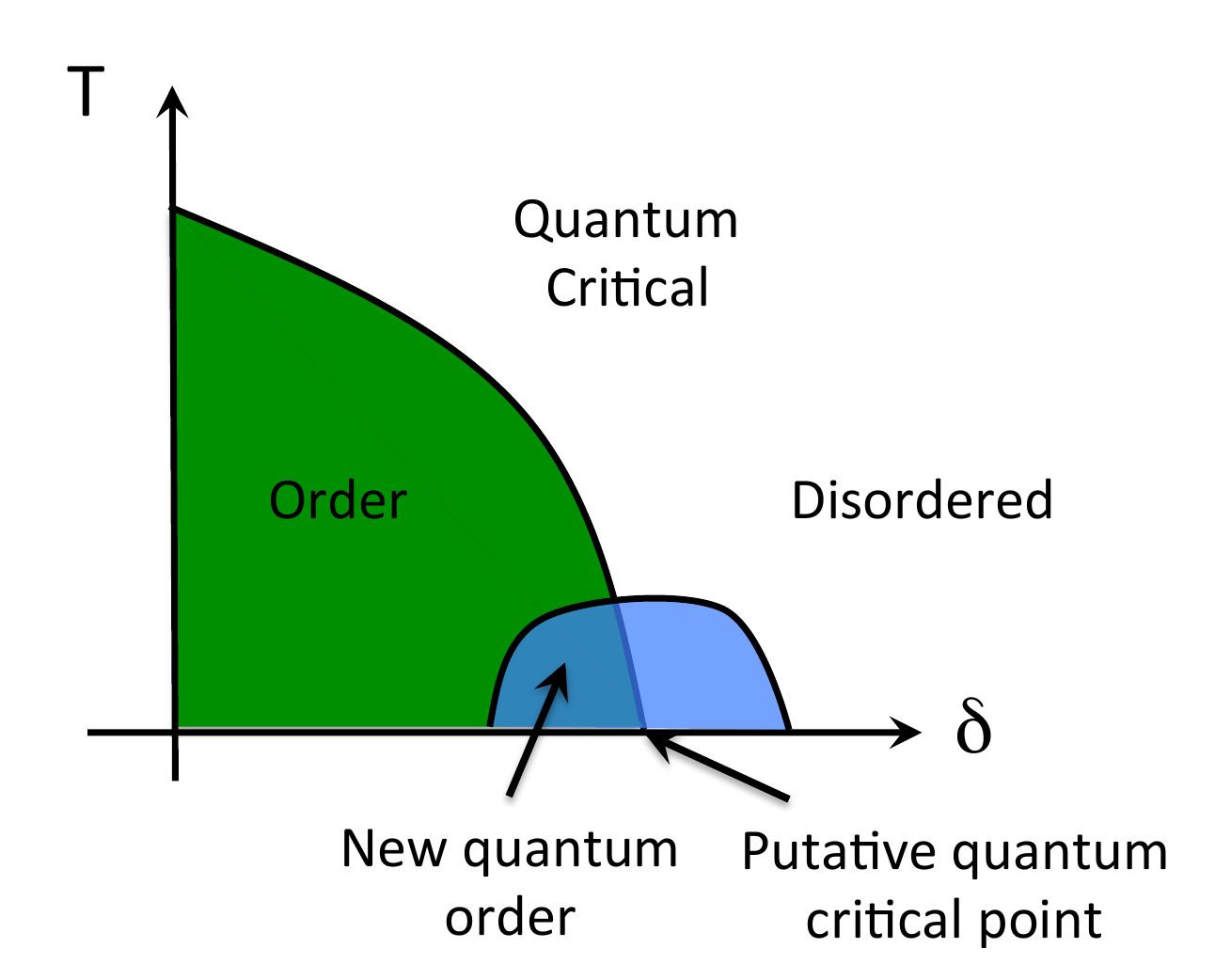}
\caption{{\bf Reconstruction of a quantum critical phase diagram:} Metallic quantum critical systems seem to be generically unstable to the formation of new phases. Perhaps gauge-gravity duality can help guide us to a principle that underlies this.}
\label{fig:QCConundrum}
\end{figure}

The thrust of this line of reasoning is in close accord with some broad themes in the study of strongly correlated electron systems. It is a matter of experimental observation that metallic quantum critical points seem to always be unstable to the formation of new phases --- when approached sufficiently closely in temperature and system parameters, new physics intervenes before criticality. This behaviour is so ubiquitous that it has led many to suspect that it is underpinned by a general principle\cite{QCConundrum} \figref{fig:QCConundrum}. The entropic drive for the holographic superconducting instability has the appeal of something quite generic. It has been found to drive other instabilities including the formation of striped phases\cite{GauntletStripes}. It also bears an amusing parallel with a relatively new, quantum order-by-disorder approach to understanding the instability of quantum critical metals to the formation of new phases\cite{Conduit:2009jt,Kruger:2012hp}. 

{\it The Fermi Surface and the Electron Star:}
An important point that has been missing from our discussion of holographic Fermi-liquids so far is the Fermi surface. This has proven to be one of the trickiest issues in the modelling of critical metals using gravitational duals. There has been much progress, which continues almost daily. 

The issue in early calculations was of a surfeit of Fermi surfaces, which show up as sharp signatures in calculated spectral functions. Essentially, there were different Fermi surfaces corresponding to the contributions from different energy scales and so distances from the black hole horizon. Several ways to resolve this issue have been discussed. One promising way to get rid of the troublesome additional Fermi surfaces 
 involves the instability of the horizon to forming an electron star.  Essentially, the charged black hole  accumulates  a cloud of charged matter near but outside of the horizon. Crucially -- and unlike the case of the superconducting instability -- a careful consideration of the interplay between the metric and the fermionic fluid seems to show that the metric is truncated, or capped off,  in a non-singular way\cite{Kachru:2008ug,Cubrovic:2011wu,HartnollStellarSpectroscopy,Sachdev:2011lx} [see Fig\figref{fig:ElectronStarandHolographicSuperconductor}]. The result is that signatures of just one Fermi surface remain.

{\it To the present:} Progress in this field continues at a feverish pace and recent developments are too numerous - and sometimes poorly digested - to list here. Active fields include studying different background metrics modelling different metallic phenomenology and the back reaction of bulk matter on the metric. One possible  interpretation of these instabilities of the charged black hole  is  that as temperature is lowered -- corresponding to approaching the horizon -- eventually marginal Fermi liquid behaviour gives way to some other behaviour. It suggests that the reason that it has proven so hard to find a Hamiltonian with a marginal Fermi liquid ground state is that one does not exist and that the marginal Fermi liquid is simply a cross-over phenomenon. There is mounting circumstantial evidence that the phenomenology that it gives way to in the gauge-gravity dual may be a form of fractionalised Fermi liquid\cite{Senthil:2003wl}.

This approach to condensed matter physics is very different from the {\it ab initio} ideal. We encode the emergent invariances of the quantum critical system in the symmetries of a background metric and use the new perspective afforded to determine the phenomenological consequences. Embedding this construction in a string theory guaranties self-consistency due to the gauge string-duality --- at least that is the hope. The aim is to deduce sum rules  of general validity and possibly new types of order.

\subsubsection{Entanglement}

No review of the gauge-gravity duality in condensed matter physics would be complete without discussing another early and very beautiful application of the approach --- the calculation of entanglement\cite{Ryu:2006za,Nishioka:2009lz}. This 
 cuts right to the underlying essence of the holographic principle. Indeed, it may be used to construct a very different perspective on the whole  construction that I will turn to in the final section. 

Entanglement has become a very important theoretical tool in the study of novel quantum many body systems. Building upon ideas from quantum information, it provides a measure of the amount of quantum non-locality in a state and has proved particularly useful in revealing topological order in the ground states of novel Hamiltonians. The general expression for the entanglement entropy is given as follows:  a system with total density matrix $\hat \rho$, is partitioned into two sub-systems A and B. The reduced density matrix for the region A is given by integrating out the degrees of freedom in region B, {\it i.e.} $\hat \rho_A=Tr_B \hat \rho$. The entanglement entropy is then given by 
\begin{equation}
{\cal S}_E=-Tr_A \left[ \hat \rho_A \ln \hat \rho_A \right].
\label{Entanglement}
\end{equation}
This expression is deceptively simple to write down, but can be notoriously difficult to calculate. 

\begin{figure}[ht]
\includegraphics[width=0.7 \linewidth]{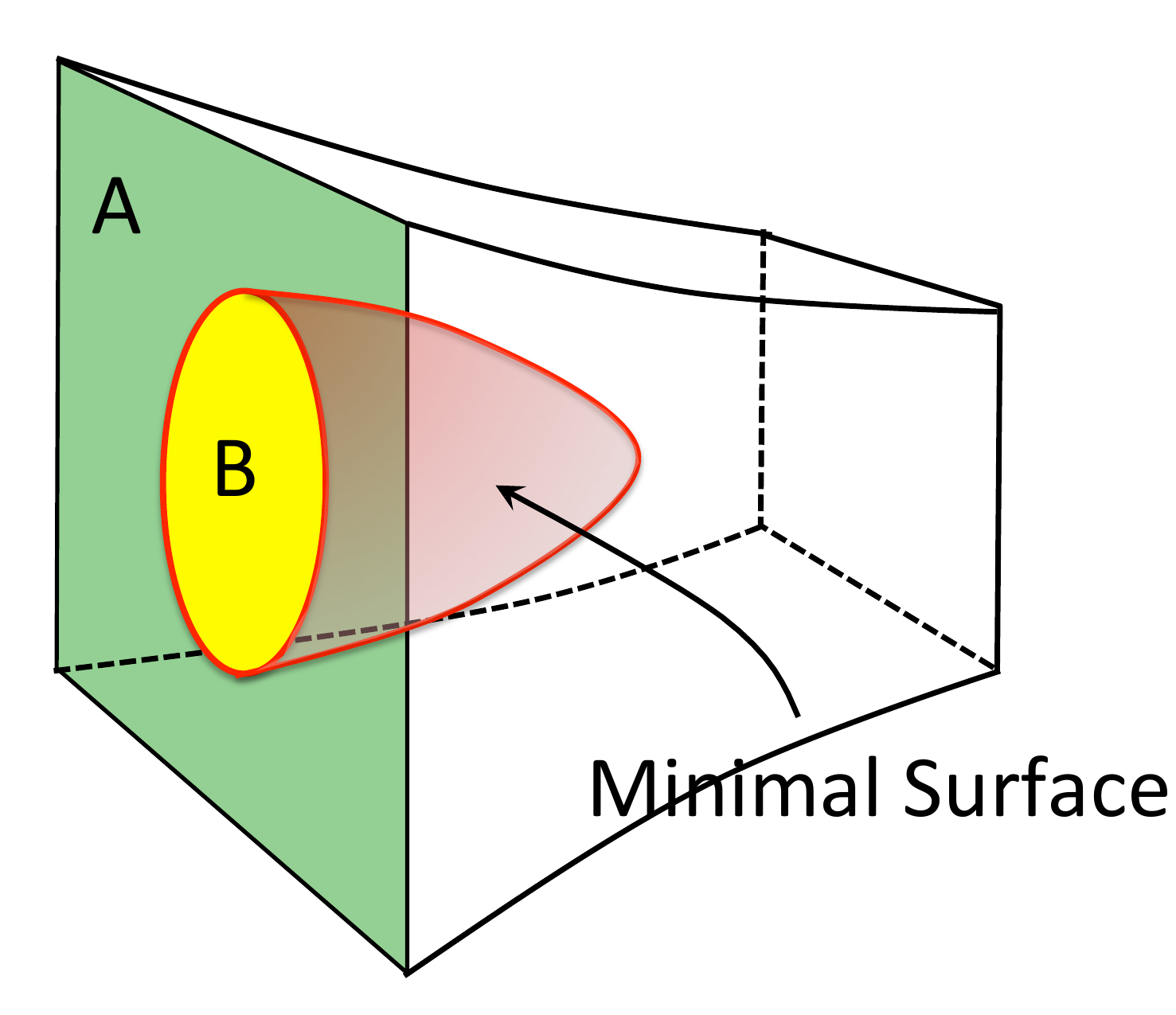}
\caption{{\bf Entanglement Entropy as a Minimal Surface:} The entanglement entropy of region B can be calculated as the area of the minimal surface extending into the bulk of the gravitational dual.}
\label{fig:Entanglemententropy}
\end{figure}

The holographic approach reduces this to the evaluation of minimal surfaces. In doing so, it draws upon the fundamental notion that the amount of information (or entropy) in a region of space-time is given by the area of the surface that bounds it. This underpins the idea of black hole entropy and indeed the holographic principle itself. The calculations of the entanglement entropy Eq.(\ref{Entanglement}) carried out by Ryu and Takayanagi\cite{Ryu:2006za} used this as follows: the theory for which we wish to calculate the entanglement entropy is put on the boundary of a higher dimensional space using the gauge gravity duality [see \figref{fig:Entanglemententropy}]. The partition into regions A and B is carried out on this boundary. A closed surface is then extended from the region A into the bulk of the extended space-time. Calculating the entanglement entropy amounts to finding the surface with the minimal area. If there is a situation where the gauge-gravity duality has simplified matters to date, this is it. 

\subsubsection{Out of Equilibrium}

Few general principles are known that govern the out-of-equilibrium behaviour of quantum systems. Those that are known constrain the system's behaviour much less than those of equilibrium statistical mechanics\cite{PhysRevLett.78.2690,PhysRevE.60.2721}. However, recent progress has shown that spatial and temporal scaling of  quantum critical and conformal system determines their out-of-equilibrium behaviour more strongly, with consequences that it might be possible to generalise\cite{PhysRevLett.93.027004,PhysRevLett.95.267001,CardyCalabrese}. The gravitational duals of such systems can be used to gain new perspective and insights into this out of equilibrium behaviour.

Three types of out of equilibrium behaviour are typically studied in condensed matter: quenches, sweeps and out-of-equilibrium steady states. To date, both quenches and non-equilibrium steady states have been studied within gauge-gravity duality. 

\vspace{0.1in}
{\it Quenches:} In quenches, the dynamics of a system is followed after a sudden change of boundary conditions. In closed quantum systems, a lot of attention has focused upon the evolution of a typical starting wave function to one in which the expectations of typical operators are the same as one would find in a thermal ensemble. In the case of the gauge gravity duality, the evolution is always one of thermal equilibration. There are a couple of ways to rationalise this that suggest some important features of the gauge-gravity duality for out-of-equilibrium systems. 

The existence of a classical metric and in particular of a horizon, implies coarse graining over some underlying quantum string variables in the gravitational dual. One may view the coarse-graining to be over an underlying pure state or, alternatively, over a thermal ensemble of underlying string states. In fact, string theorist would not usually think of an ensemble, but it reasonable on the condensed matter side, since we typically do have a statistical ensemble rather than a pure state. For the condensed matter physicist, therefore we can think of the black hole horizon as providing a thermal bath with which the rest of the system may equilibrate. It is not surprising then that the system will tend to a thermal state after a quench. 

A subtlety in this picture concerns the role of the large number of colours in the string embedding of the gauge-gravity duality. This large number of colours suppresses string fluctuations and guarantees the existence of a classical metric. It is not yet fully clear what constraints this places on the types of system whose out-of-equilibrium dynamics can be studied by holographic methods. 

Nevertheless, the gauge gravity duality has revealed some very interesting details about how the thermal distribution is approached after a quench. This follows from the understanding of black hole formation. A small number of modes determine the late-stage dynamics of this process and so too a small number of quasi-normal modes will determine thermal equilibration. 

These ideas have been developed to study the behaviour of a metal when a Cooper interaction is suddenly switched on. Original studies of this process in the condensed matter literature showed an oscillation of the superconducting order parameter in the closed system, which is expected to decay in time for the open system\cite{Barankov:2004em}. This physics is reproduced in the gravitational dual with the possibility of drawing some rather general conclusions\cite{Bhaseen:2012ki}. 

\vspace{0.1in}
{\it Out-of-Equilibrium Steady State:} Out-of-equilibrium steady states are largely determined by a system's dynamics. In the case of quantum critical systems, one might then anticipate that universal dynamics will result in universal out-of-equilibrium steady states. This notion has been proved correct in a small, but growing number of cases\cite{PhysRevLett.97.236808,PhysRevB.78.195104}. The results of such analyses for the Bose-Hubbard model is
of particular note in the context of gauge-gravity duality, since the calculation of gauge fluctuations in the gravitational dual seem to be in such close accord with the equilibrium properties of the Bose-Hubbard model. 

The steady-state equilibrium behaviour of the Bose-Hubbard model in the presence of an electric field was studied originally by Green and Sondhi\cite{PhysRevLett.95.267001} who showed how fluctuations created by Landau-Zener breakdown of the vacuum ({\it i.e.} dielectric breakdown or the Schwinger mechanism) could in certain circumstances undergo critical scattering leading to the formation of a universal out of equilibrium steady state. Although the novel features of this steady state were not immediately apparent in the steady state current, later analysis showed that the current noise provides a revealing window upon it\cite{PhysRevLett.97.227003}.

A gravitational dual of these effects was first constructed by Karch and Sondhi\cite{Karch:2010kt}. They considered a model embedded in an 9+1-dimensional space with a background metric produced by many planar 3-dimensional objects  known as 3-branes --- or black branes, since in this context they lead to the formation of an event horizon. The non-equilibrium steady state was investigated by calculating the properties of gauge fields on a 5-dimensional surface (a so-called D5 brane) that intersects this D3-brane. Gauge fields in the bulk are dual to currents in the 2-dimensional boundary of this space. Their properties were investigated in the presence of boundary conditions corresponding the application of a uniform electric field. These boundary conditions modify the metric induced upon the D5-brane such that the horizon is characterised by an effective temperature $T_{eff}$ given by a combination of the background temperature $T$ and the applied electric field, $E$;
$$ T_{eff}=[T^4+E^2]^{1/4}.$$
The steady-state current found in this way recovers the results of Ref.\cite{PhysRevLett.95.267001}.
An extension to this work showed that Hawking radiation produced at the horizon and propagating to the boundary is dual to current fluctuations on the boundary\cite{Sonner:2012dz}. This reproduced previously reported results for the out-of-equilibrium current noise and extended them to interpolate -- in a universal way through the form of the new effective temperature -- between low- and high-field limits. 

\begin{figure}[ht]
\includegraphics[width=0.7\linewidth]{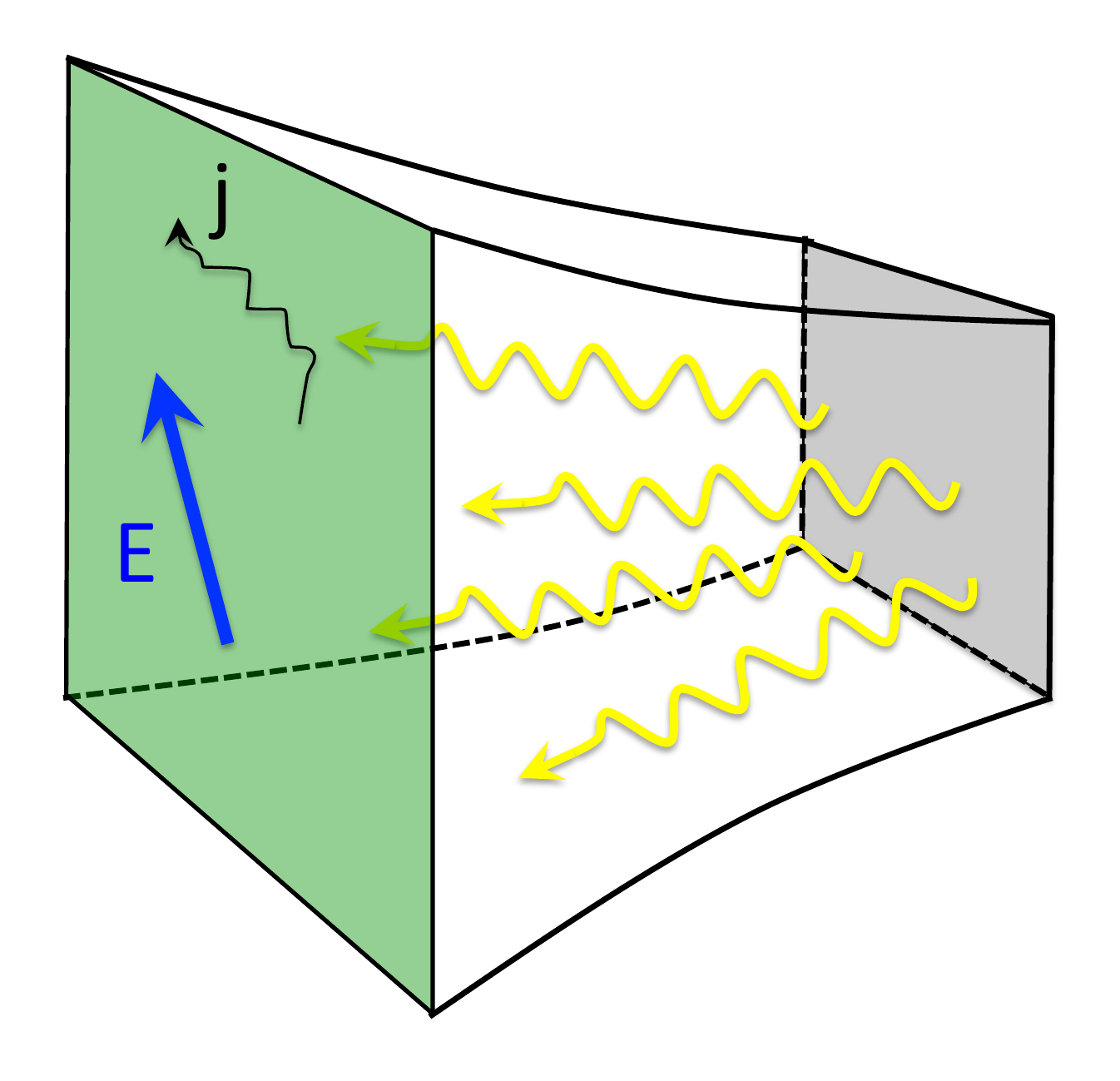}
\caption{{\bf Hawking Radiation is Dual to Current Noise in the Boundary Theory}}
\label{fig:HawkingvsCurrentNoise}
\end{figure}

The solution has a number of interesting features the consequences of which have not all be fully worked out. In general, a driven, non-equilibrium steady state need not be thermal. Remarkably, the current fluctuations found in the gravitational dual are guaranteed to have a thermal distribution. This is a consequence of the static background metric --- the Hawking radiation emanating from the horizon has a thermal distribution and the distribution remains thermal upon propagation to the boundary [see \figref{fig:HawkingvsCurrentNoise}]. Such effects have been seen in out-of-equilibrium analyses of quantum critical systems from the condensed matter perspective, but it is not clear yet if some unifying principle underlies it. The gravitational duality suggests that, if this result is correct, it is indeed the result of a deep underlying principle. 

However, there is reason for caution. In order to obtain a universal steady state on the condensed matter side, one must think very carefully about heat flow\cite{PhysRevA.19.1721}. Joule heat generated in the bulk must be transported to the boundary. If the resulting steady state is to be universal and independent of the shape and geometry, the transport of heat must be very rapid and the rate-limiting step for equilibration should be the scattering of heat into these thermal transport modes. Both the original condensed matter analysis and the gravitational duals circumvent these difficulties by cunning analytical tricks. In the gravitational dual, this amounts to the probe limit whereby there are very many more D3-branes than D5-branes. This reduces the feedback effect of the non-equilibrium gauge fields upon the total, 10-dimensional metric and effectively demands that the final distribution of fluctuations is thermal. It is not clear yet to what extent this is an artefact of the set up and more investigation is required.

\section{Perspective and Future Prospects}

The results described above were obtained assuming the validity of the gauge-gravity duality described by Eq.(\ref{Duality}). Despite the very good reasons for believing this validity, several assumptions underlie these analyses. The supersymmetric string theory in which the gauge-gravity duality was derived does not correspond to any condensed matter Hamiltonian and, moreover, the precise role of the many colour (large $N$) limit  is not yet clear.   Indeed the situation is even more delicate, since at present we cannot deduce the gravitational dual corresponding to any particular microscopic condensed matter Hamiltonian.  As emphasised in the preceding discussion, the spirit of these calculations is to seek universal constraints upon the behaviour of systems falling into certain classes. 

The import of this approach for condensed matter physics is that the resulting phenomenology is guaranteed to be internally consistent by the string theoretical embedding. This can have far reaching consequences, some of which have been highlighted. The situation is, however, philosophically very different from the {\it ab initio} ideal that pervades much of condensed matter physics.

Even accepting the power of this approach, there are good reasons to attempt to construct the gravitational dual to a given condensed matter system directly. Not least amongst these is the determined independence of condensed matter physicists. It rests rather uncomfortably to take anything upon the authority of mathematical tools with which one is unfamiliar. This is a tricky task. In constructing such a dual theory, we must in a sense solve precisely the problems that we in the condensed matter community have been struggling with.  Nevertheless, there are a couple of approaches that may yet yield insights that do not directly involve the machinery of string theory.

\subsection{Gauge-Gravity Duality as a Novel RG}

Perhaps the most direct attempts to obtain a constructive derivation of the gauge gravity duality flow from the insight that the additional dimension in the gravity dual tracks the renormalization group flow\cite{Wilson:1971eb,Wilson:1971fo}. This has inspired several different approaches that aim to construct the new, emergent dimension by explicitly tracking the renormalization group flow from the the boundary theory. I will focus upon two.

{\it Momentum-shell renormalization:} The essence of the approach of Sung-Sik Lee is a modification of the momentum shell renormalization familiar in condensed mattercite{Lee:2010fk,Lee:2012fk}. The starting point is the generating function of the field theory, Eq.(\ref{GeneratingFunction}). At each stage of the renormalization, fields in a high momentum shell are integrated out. Unlike the conventional momentum-shell renormalization, these degrees of freedom are not integrated in favour of modified parameters in the theory of the remaining modes. Rather, one allows the integration of these high energy degrees of freedom to generate dynamics for the source fields.

A variable tracking how far along the renormalization group trajectory one has progressed forms the new, emergent dimension of the gravity dual. In this way, we arrive at a theory with an additional dimension in which the source fields of the boundary generating function have become dynamical in the bulk ({\it i.e. } new terms generated in the Hamiltonian imbue them with dynamics) with boundary conditions given by the source fields. This is precisely the content of Eq.(\ref{Duality}). 

Perhaps, with hindsight,  this geometrization of the energy scale could have been anticipated, since the equations of the renormalization group are local differential equations. However,
attempts to go through this construction for an arbitrary boundary theory lead to something very non-local in the bulk of the extended space. Only very special theories lead to a sensible bulk theory --- those  with quantum critical scaling. At the very least, this shows the powerful and far-reaching consequences of a quantum critical scaling ansatz. 

Whether such an approach will eventually lead to a constructive route to gravitational duals of condensed matter Hamiltonians is not yet clear. There are good reasons for caution. The calculations may be difficult in practice --- after all, the fixed point phenomenology of a given condensed matter Hamiltonian is essentially the problem that we have been trying to solve all along. Nevertheless, its existence even  in principle, together with the sum rules and constants obtained from the gravitational dual, can perhaps pave the way for new non-perturbative and self-consistent treatments of the condensed matter Hamiltonian. 


{\it Real space renormalization:} Ideas from quantum information theory have had a significant impact upon many-body quantum mechanics in recent years. In particular, the realisation of the central role played by entanglement in determining the properties of quantum states has been an important boon. 

\begin{figure}[ht]
\hspace{-0.8 \linewidth}
a)
\\
\includegraphics[width=1.\linewidth]{{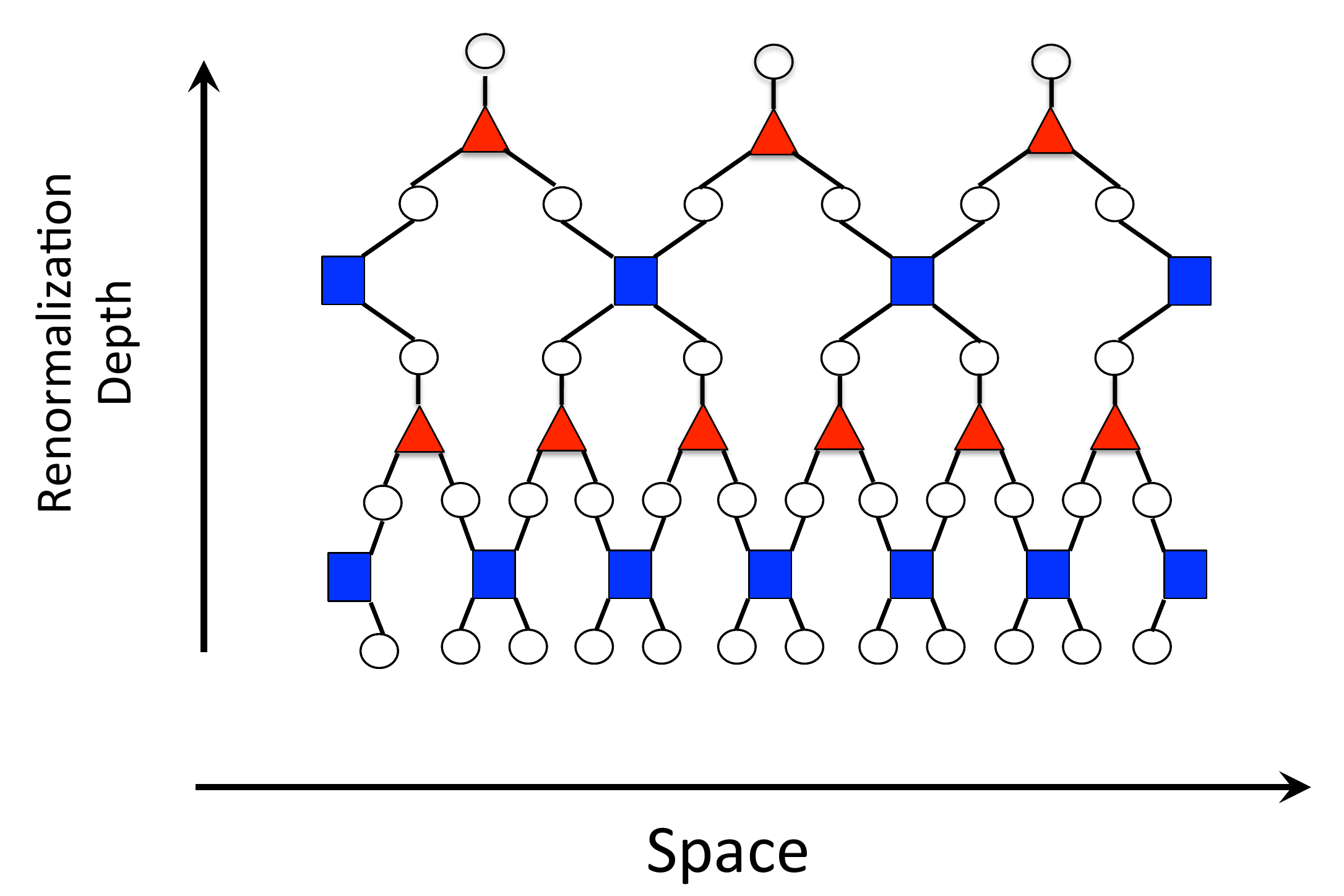}}
\\
\hspace{-0.8 \linewidth}
b)
\\
\includegraphics[width=0.7 \linewidth]{{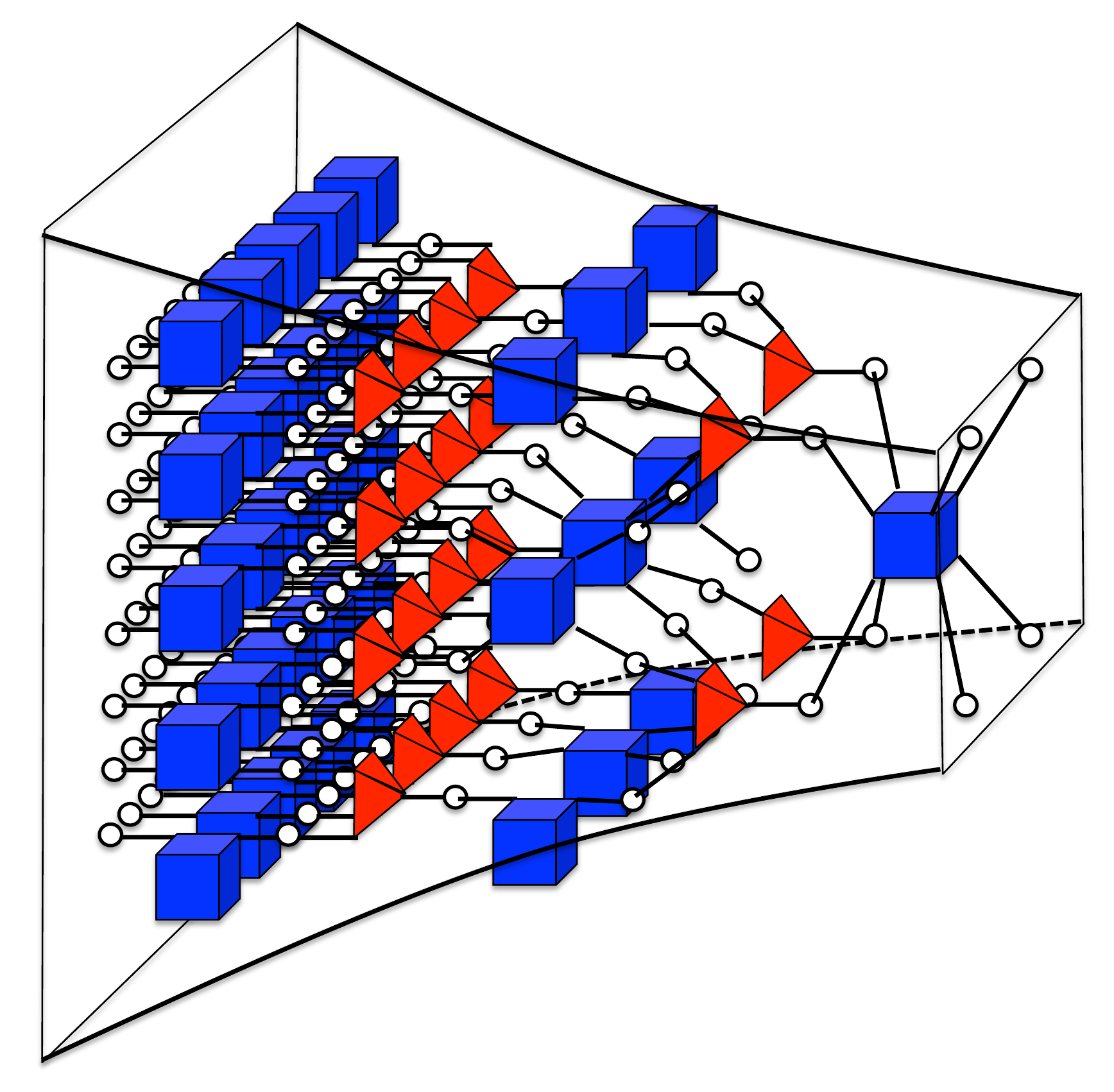}}
\caption{{\bf The Multi-Scale Entanglement Renormalization Ansatz (MERA):}  a) A schematic representation of the MERA. Layers of the ansatz wave function are produced by a block renormalization proceeding in two steps i. {\it Disentangling}  - square sumbols - a unitary transformation is used to disentangle along the boundary of the block, ii. {\it Coarse-graining (Isometry)} - triangular symbols - whereby the degrees of freedom are reduced whilst keeping the entangled contributions to the density matrix. b) This is a microscopic construction of a holographic duality.}
\label{fig:MERA}
\end{figure}

The multi-scale entanglement renormalization ansatz (MERA) is a method that has recently been developed to provide improvements in the numerical modelling of lattice Hamiltonians\cite{Vidal:2007zv,Vidal:2008pz}. It can be thought of as a development of the density matrix renormalization group (DMRG)\cite{White:1992hz} focussed upon preserving entanglement. While the DMRG focusses upon optimising a matrix product state variational wavefunction for a given system, the MERA optimises a variational ansatz with a novel layered tensor network structure in which layers of increasing depth posses entanglement on increasing length scales. The structure of this ansatz is illustrated in \figref{fig:MERA}. 

As emphasised by Vidal\cite{Vidal:2007zv,Vidal:2008pz}, the use of this ansatz may be interpreted as an unusual renormalization group flow. Layers at different depths of the tensor wave function have Hamiltonians associated with them that scale in a particular way. The additional dimension introduced by the multi-scale entanglement ansatz plays  a role essentially identical to the additional dimension of the gravity dual\cite{Swingle:2012hw}. Indeed the Ryu-Takayanagi\cite{Ryu:2006za} construction of the entanglement entropy of a region in terms of minimal surfaces has an appealing parallel in MERA. The details of how the multi scale entanglement ansatz can be interpreted as a discretized version of gauge-gravity duality are given in Ref.\cite{Swingle:2012hw}.

Although apparently philosophically very different from the momentum-shell renormalization group construction, focussing upon entanglement leads to a very similar route to the gravitational dual. It is notable that the MERA was developed in seeking more rapid and accurate ways to determine {\it numerical} solutions for complicated many-body problems. This is a salutary message in condensed matter physics. New insights can arise from addressing the difficult task of actually performing calculations. In a similar vein, the central role of the nodal surface in the density matrices of fermonic systems was revealed in attempts to perform quantum Monte Carlo simulations\cite{CeperleyNodes}. Some are pursuing this as an alternative route to gauge-gravity duality\cite{Kruger:2008bw}.

\section{Conclusions}

The ragged edge of theoretical physics is riven with mathematical inconsistency. Resolving these can often lead to new insights. More commonly, perhaps, incorporating new facts as they are thrown up by experiment can also iron out these inconsistencies (only to replace them with others). In the late 1970s, a dearth of new data in the right area meant that those trying to formulate fundamental theories of the universe were left with only their mathematical imagination to drive progress.

This spawned a tremendous period of creativity underpinned by the hope that a theory which encompassed all of the known physical laws in a mathematically consistent way would provide both a complete and unique description of the universe. String theory was the result. Unfortunately, it is not clear whether or not there is a unique translation of these ideas to the 4-dimensional space-time of our experience. 

Still, string theory comprises a powerful new set of mathematical tools. Some of the most profound insights that have stemmed from it have been in mathematics and geometry. At the same time, attempts to develop unified theories starting from a range or different physical motivations now use essentially these mathematical ideas. In this sense, any fundamental theory will now look like a string theory as it will inevitably use these tools.

Perhaps, then, it is inevitable too that we should turn these new mathematical tools to the vexing problems of condensed matter physics. The field has been coming pretty close to these ideas in any case. The link is, in itself, something remarkable and worthy of celebration. We are used to the idea of an intellectual thread that runs through science from particle and nuclear physics, through chemistry and biology, and ultimately to astronomy and cosmology. The gauge gravity duality shows that this thread is more ramified than hitherto anticipated. It exposes a unity in theoretical physics that was previously clear only to the most insightful amongst us.

There are good reasons, too, for optimism about the utility of gauge-gravity duality. Recasting the fundamental principles and questions of physics under such contrasting lights has already led to new insights and results. More will doubtless follow. This is not the only game in town for the condensed matter physicist. Alongside analytical advances, new numerical tools such as dynamical mean field theory, density matrix renormalization group and steady advances in quantum Monte Carlo have brought progress to problems that have eluded analytical treatment. These certainly have a leading role to play and, as described above in the case of MERA, can lead to dramatic insights. 
This is an exciting time for physicists working in quantum many-body theory. Despite the tendency of human organisations to balkanise,
science periodically forces us to accept the unity of our endeavour. This is one such juncture.

{\it Acknowledgments:} I would like to thank Joe Bhaseen, Julian Sonner and Jonathan Keeling  for their critical reading of this manuscript and the EPSRC for funding  under grant code EP/I004831/1.

\end{document}